\documentclass[useAMS,usenatbib,fleqn]{mnras}
\usepackage{newtxtext,newtxmath}
\usepackage[T1]{fontenc}
\usepackage{ae,aecompl}
\usepackage{graphicx}   
\usepackage{amsmath}    
\usepackage{amssymb,times}    

\def\pz{PZ~Mon}
\def\rs{RS~CVn}
\def\Teff{$T_{eff}$}
\def\Tspot{$T_{s}$}
\def\Twarm{$T_{w}$}
\def\Sspot{$S_{s}$}
\def\Swarm{$S_{w}$}

\def\lgg{log~$g$}

\def\bv{\hbox{$B\!-\!V$}}
\def\ebv{E(\bv)}
\def\deg{^\circ}

\def\actaa{Acta Astron.}      

\begin{document}

\title[Cool spots on the surface of the active giant \pz]{Cool spots on the surface of the active giant \pz}

\author[Yu.V.Pakhomov et al.]{
\parbox{\textwidth}{
Yu.~V.~Pakhomov$^{1}$,
V.~I.~Shenavrin$^{2}$,
N.~I.~Bondar'$^{3}$, 
K.~A.~Antonyuk$^{3}$, 
N.~V.~Pit'$^{3}$,
S.~P.~Belan$^{3}$,
S.~Yu.~Gorda$^{4}$
}\\
\\
$^1$Institute of Astronomy, Russian Academy of Sciences, Pyatnitskaya 48, 119017, Moscow, Russia\\
$^2$Sternberg Astronomical Institute, Moscow State University, Universitetskii pr. 13, Moscow, 119992 Russia\\
$^3$Crimean Astrophysical Observatory RAS, Nauchny, Republic of Crimea, Russia\\
$^4$Kourovka Astronomical Observatory, Ural Federal University, ul. Mira 19, Yekaterinburg, 620002 Russia
}

\date{\noindent
ISSN 1063-7737, DOI: 10.1134/S1063773719030058\\
Astronomy Letters, 2019, Vol. 45, No. 3, pp. 156--163.
}

\pagerange{\pageref{firstpage}--\pageref{lastpage}}
\pubyear{}

\maketitle
\label{firstpage}

\begin{abstract}
Based on the multiband ($BVRIJHKL$) photometric observations of the active red giant
PZ Mon performed for the first time in the winter season of 2017--2018, we have determined the main
characteristics of the spotted stellar surface in a parametric three-spot model. The unspotted surface
temperature is \Teff = 4730~K, the temperature of the cool spots is \Tspot = 3500~K, their relative area is about
41\%, and the temperature of the warm spots is \Twarm$\approx$4500~K with a maximum relative area up to 20\%. The
distribution of spots over the stellar surface has been modeled. The warm spots have been found to be
distributed at various longitudes in the hemisphere on the side of the secondary component and are most
likely a result of its influence.

\end{abstract}

\begin{keywords} 
stars: individual: \pz\ --
(stars:) starspots --
stars: variables: general
\end{keywords}

\section*{Introduction}
\noindent
The red giant \pz\ (HD\,289114, $V\approx9.3$~mag, K2III) is an RS CVn variable star (a period of 34.13 days) with a pronounced activity \citep{2015MNRAS.446...56P} and a low-mass component synchronously revolving in a circular orbit \citep{2015AstL...41..677P}. The activity of stars of this type manifests itself as variability of the brightness modulated with their rotation period and is usually explained by the interaction with the gravitational and magnetic fields and the radiation field of the companion, which affect the structure of the atmosphere and magnetic field of the primary component. At the locations where the magnetic field lines emerge the convective energy transfer is suppressed, which leads to some surface cooling relative to the quiet photosphere and the appearance of cool temperature spots. The amplitude of brightness variations depends on the physical characteristics of the stars in the binary system. According to the catalogue by \citep{2008MNRAS.389.1722E}, significant brightness differences, up to 0.3$^m$, are observed in the case of close stars with a mass ratio $q=M_2/M_1>0.3$ and in the case of a hot companion. Doppler tomography is applied to study the magnetic field of stars. This method requires polarimetric spectroscopic observations with a high resolution and a high signal-to-noise ratio, which restricts considerably the number of accessible objects. An analysis of photometric light curves allows the surface structure and evolution of cool temperature spots to be studied qualitatively for a large number of objects, including faint ones. The component mass ratio in the binary system PZ Mon, $q=0.09$, is the minimum one among the known synchronous RS CVn stars, while the brightness amplitude is significant, up to 0.10-0.15$^m$ in the $B$ and $V$ bands. At the same time, the active region is fairly stable and is always located on the side of the secondary component \citep{2015AstL...41..677P, 2017ASPC..510..128P}. This determines the interest in studying the nature of the activity of PZ Mon and the extent to which it is affected by the secondary component.

From October 2017 to April 2018 photometric observations of PZ Mon were performed for the first time in a wide spectral range, from the optical one in the B band to the infrared in the L band. The goal of this paper is to determine the photospheric characteristics of the active giant PZ Mon at the epoch of observations by analyzing and modeling its light curves. In Section 1 we describe our photometric observations. In Section 2 we analyze the maxima and amplitudes of the light curves and estimate the parameters of cool and warm spots. In Section 3 we construct a model for the distribution of spots over the surface of PZ Mon that describes its photometric characteristics. Next we present a discussion of our results and our conclusions.

\section{OBSERVATIONS}
\label{sec-observations}
\begin{figure}
	\centering
	\includegraphics[width=\columnwidth,clip]{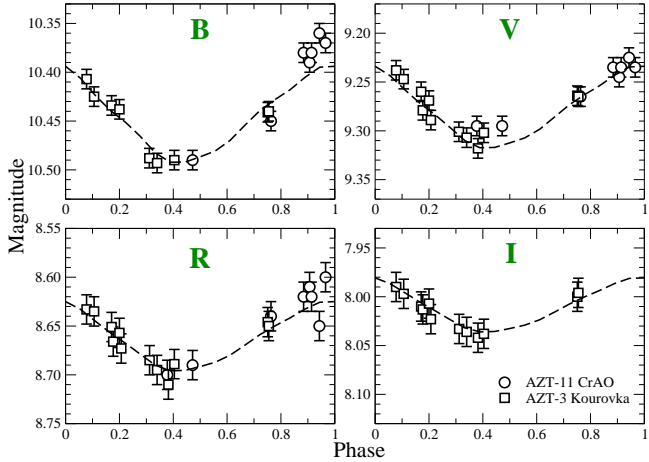}
	\caption{$BVR$$_C$$I$$_C$ light curves of \pz. The squares and circles are the data obtained at the Kourovka Astronomical Observatory and the Crimean Astrophysical Observatory, respectively. The dashed curves are the computed light curves. The scale along the magnitude axes is the same.}
	\label{fig:BVRI}
\end{figure}

\begin{figure}
	\centering
	\includegraphics[width=\columnwidth,clip]{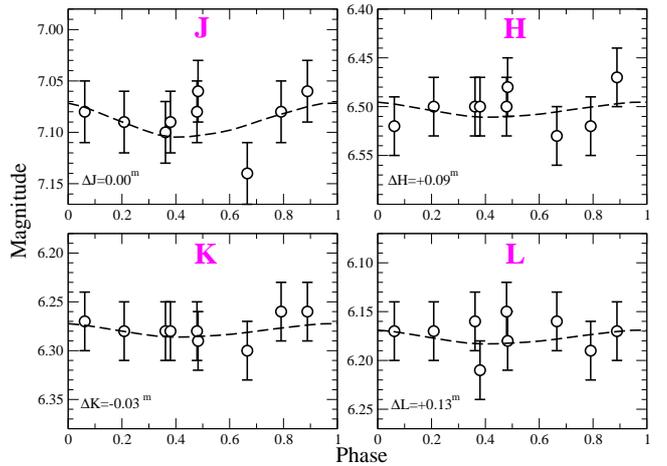}
	\caption{$JHKL$ light curves of \pz. The dashed curves are the light curves computed with a shift whose value is presented
	in the lower left corner of each plot. The scale along the magnitude axes is the same as that in Fig.~\ref{fig:BVRI}.}
	\label{fig:JHKL}
\end{figure}

\noindent
Our photometric observations of PZ Mon were performed with three instruments in the winter season of 2017--2018. From December 26, 2017, to March 7, 2018, we carried out seven observations in three bands, $BVR_C$, each of which consisted of several exposures, with the 1.25-m AZT-11 telescope of the Crimean Astrophysical Observatory using a five-channel ProLine PL230 photometer equipped with an e2v CCD230-42 CCD array. The reduction to the standard Johnson--Cousins photometric system was based on the previously determined transformation coefficients. The accuracy of our magnitude measurements ranged from 0.007$^m$ to 0.010$^m$. From January 5 to April 3, 2018, we performed 12 observations in four bands, $BVR_CI_C$, with the 0.45-m AZT-3 telescope of the Kourovka Astronomical Observatory, the Ural Federal University, using an FLI PL230 CCD camera (e2v CCD230-42-1-143, 2048$\times$2048, 15~$\mu$ pixel size). The reduction to the standard Johnson-Cousins photometric system was made using stars in the same frame based on data from the APASS catalogue \citep{2015AAS...22533616H}. The accuracy of our magnitude measurements ranged from 0.008m to 0.015$^m$. From October 11, 2017, to February 15, 2018, we carried out nine observations in four bands, $JHKL$, with the 1.25-m ZTE telescope at the Crimean astronomical station of the Sternberg Astronomical Institute, the Moscow State University, using an infrared photometer based on an InSb photovoltaic detector. The accuracy of our magnitude measurements ranged from 0.02$^m$ to 0.03$^m$. The total time of our observations covers about five axial rotation periods of PZ Mon, the bulk of which are within two periods. According to the ASAS-SN photometry \citep{1997AcA....47..467P}\footnote{https://asas-sn.osu.edu/variables}, there were no changes in the brightness maxima and amplitude at the epoch under study. Although the accuracy of ASAS magnitude measurements is relatively low (0.02--0.05$^m$), such a stable behavior is typical for the star under study and, hence, it is possible to investigate the light curve of PZ Mon folded with the previously found period. The epoch for zero phase was taken to be JD~2454807.2, corresponding to the brightness maximum. Multicolor photometry for PZ Mon in such a wide range of bands has been obtained for the first time. Figures~\ref{fig:BVRI} and~\ref{fig:JHKL} show the light curves for the optical and infrared ranges, respectively. For clarity, the scale along the magnitude axis was kept the same for all the light curves.

\section{PHOTOMETRIC CHARACTERISTICS OF \pz}
\label{sec:photometry}
\begin{figure}
	\centering
	\includegraphics[width=\columnwidth,clip]{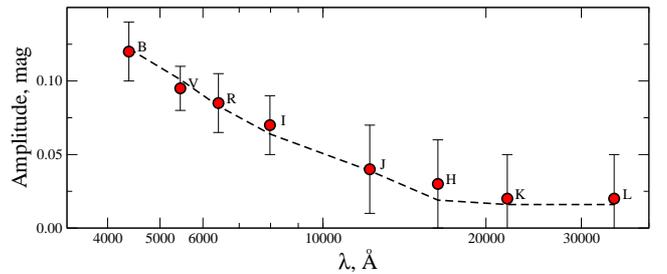}
	\caption{Observed brightness amplitude (circles) versus effective wavelength of the bands. The dashed curve indicates the computed amplitudes that describe best the observations.}
	\label{fig:amp}
\end{figure}

\noindent
The same behavior of the light curves and a gradual decrease in the amplitude with increasing effective wavelength of the band can be seen in Fig.~\ref{fig:BVRI} and~\ref{fig:JHKL} drawn on the same scale. The largest amplitude (Fig.~\ref{fig:amp}) is typical for the short-wavelength $B$ and $V$ bands, while the amplitudes for the infrared bands are comparable to or smaller than the observational errors; the IR brightness hardly exhibits any variability.

\subsection{Coll spot}
\noindent

\begin{figure}
	\centering
	\includegraphics[width=\columnwidth,clip]{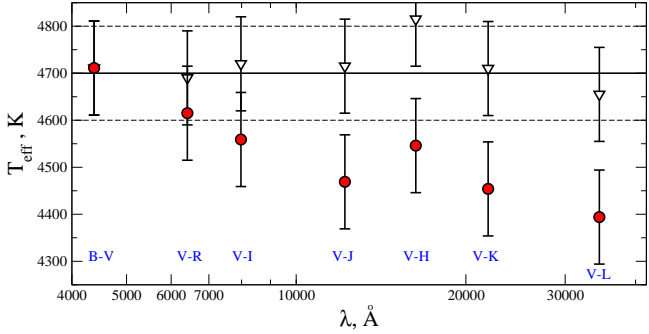}
	\caption{The effective temperatures of \pz\ determined from the observed colors at maximum light (circles) and the reconstructed colors (triangles). The straight line is the value of Teff estimated from our spectroscopic observations with the error indicated by the dashed lines.}
	\label{fig:T}
\end{figure}

\begin{figure}
	\centering
	\includegraphics[width=\columnwidth,clip]{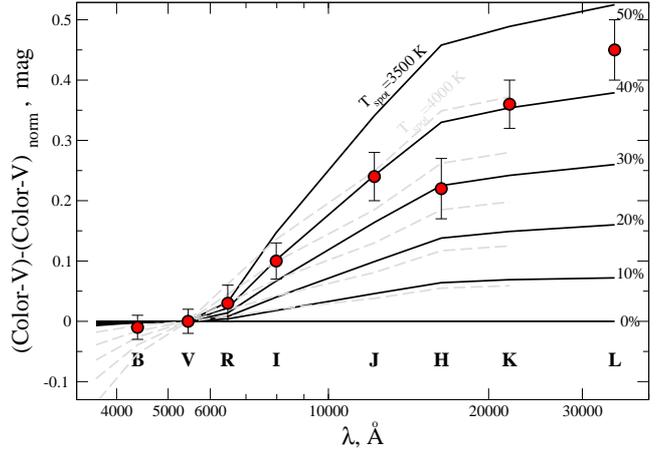}
	\caption{Color excesses of \pz\ relative to a normal star with parameters \Teff/\lgg/[Fe/H]/\ebv = 4730/2.8/0.07/0.06: the circles are the observed values for the epoch of \pz\ observations; the solid lines indicate the theoretical values computed for a star with a spot temperature \Tspot=3500~K and a relative spot area $S$ from 0 to 50\%; the dashed lines indicate the theoretical values computed for a star with a spot temperature \Tspot=4000~K.}
	\label{fig:dm}
\end{figure}

The colors derived from $BVRIJHKL$ observations depend on the star's effective temperature and are used for its estimation. Figure~\ref{fig:T} shows the effective temperatures of PZ Mon determined from various colors based on synthetic spectra \citep{1998A&A...333..231B}. \Teff\ for the \bv\ color is close to its value obtained by analyzing optical spectral lines \citep{2015MNRAS.446...56P}, while a different picture is observed for other colors: Teff decreases with increasing effective wavelength of the band. On average, the values of the \bv\ color itself lie in the range from 1.16$^m$ to 1.18$^m$, which has been observed already for a long time even at considerable brightness $V$ differences from 9.0$^m$ to 9.5$^m$. These observational facts can be interpreted as the presence of a cool region with a considerable area in the stellar photosphere \citep{2018AstL...44...35P}. Indeed, a surface considerably cooler than the quiet photosphere emits much less energy in the visible range, for example, in the $B$ band: $F(4730 K)/F(3500 K) \approx 20$; therefore, its presence is not revealed by an analysis of the short-wavelength part of the optical range. The cool region is much more pronounced in the long-wavelength part of the spectrum (in the $K$ band: $F(4730 K)/F(3500 K) \approx 1.5$), as a result of which the temperature determined from increased IR colors will be considerably underestimated. Figure~\ref{fig:dm} shows the color excesses at the brightness maximum of PZ Mon relative to a normal unspotted star with the same parameters (\Teff=4730~K, \lgg=2.8, [Fe/H]=0.07, \ebv=0.06~mag). Both the observed values and the computed ones in the presence of a spot with \Tspot=3500 and 4000~K and its area S from 0 to 50\% are displayed. The computations were carried out by interpolating the grid of magnitudes $m_X(T_{eff},\textrm{lg}\,g, [Fe/H])$ in the corresponding bands $X$ from \citep{1998A&A...333..231B}\footnote{http://wwwuser.oats.inaf.it/castelli/grids.html}.
The magnitudes of the spotted star were computed by partially taking into account the limb-darkening effect (the magnitudes $m_V$ themselves from the grid of models include this effect, but the distribution of spots over the surface is not considered at the moment) from the relation

\begin{eqnarray}
m_X = -2.5\,lg (10^{-0.4 \displaystyle m_X^s}\,S+10^{-0.4 
	\displaystyle m_X}\,(1-S)) - 
\nonumber\\
-5\, lg\frac{\theta}{2} + E(B-V) R_X
\label{eq:m}
\end{eqnarray}
\noindent
where $m_X^s$ is the magnitude of 1~cm$^2$ of the cool stellar surface at \Tspot=3500~K, $m_X$ is the magnitude of 1~cm$^2$ of the unspotted stellar surface at \Teff = 4730 K, $\theta=0.29$~mas is the angular diameter of PZ Mon, $R_X=A_X/$\ebv\ is the ratio of the extinction in the
band to the \bv\ color excess. We adopted \ebv=0.06~mag, $R_X$= 4.08, 3.10, 2.58, 1.85, 0.88, 0.56, 0.34, and 0.28 for the $BVRIJHKL$ bands,
respectively. 

\citet{2018AstL...44...35P} estimated the cool spot temperature \Tspot=3500~K by analyzing the TiO molecular band. It can also be seen from Fig.~\ref{fig:dm} that a considerable \bv\ color excess is expected at a higher temperature, \Tspot=4000~K, but this is not observed. The color excesses agree with the computed values for an area of the cool region 41$\pm$2\%\ of the visible stellar surface. We can use these values together with the observed ones and reconstruct the colors and temperature of the unspotted surface of PZ Mon. Such reconstructed temperatures for different colors are presented in Fig.~\ref{fig:T}. The mean effective temperature \Teff=4717$\pm$45~K computed using all colors is in good agreement with \Teff=4700$\pm$100~K, the temperature determined by analyzing spectral lines \citep{2015MNRAS.446...56P}, and \Teff=4730$\pm$50~K derived from our 2016--2017 photometric data \citep{2018AstL...44...35P}. In what follows, we adopt \Teff=4730~K.

When computing the magnitudes (eq.\ref{eq:m}), the effects of interstellar extinction and cool spots are similar. Both these effects exert a smaller influence in the IR part of the spectrum than in the visible one, i.e., \ebv\ and $S$ are dependent quantities, i.e., an increase in one of them leads to a decrease in the other. The previously adopted \ebv=0.06~mag was derived from the observed \bv\ color and the normal one computed for the temperature \Teff=4700~K determined by analyzing the spectrum. To estimate the influence of the accuracy of the interstellar  reddening \ebv\ on the accuracy of the spot area $S$, we approximated the magnitudes of PZ Mon at maximum light by the theoretical values (eq.\ref{eq:m}) using the Levenberg--Marquardt method. The results of our computations are $S$=38$\pm$6\%\ and \ebv=0.08$\pm$0.03, which confirm the values obtained previously within the error limits. 

\begin{figure}
	\centering
	\includegraphics[width=\columnwidth,clip]{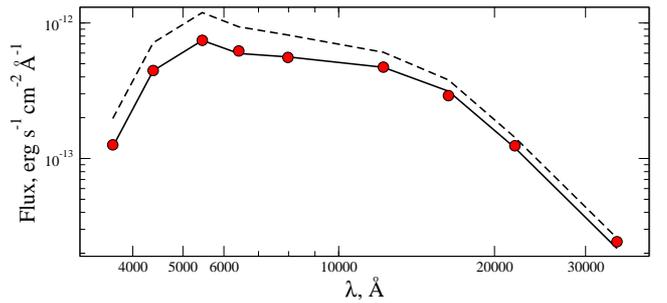}
	\caption{Spectral energy distribution for \pz\ (circles) compared to a normal star (dashed curve) with parameters \Teff/\lgg/[Fe/H]/\ebv = 4730/2.8/0.07/0.06. The solid curve is the theoretical distribution in the presence of spots with \Tspot = 3500~K and a relative area $S$ = 41\%.}
	\label{fig:sed}
\end{figure}

In the presence of such a significant cool region, significant brightness fluctuations, especially in the $B$ band, with an amplitude up to  $-2.5\times \mathrm{log}(1-0.4+0.4*1/20)=0.52^m$ are also expected. However, the brightness amplitude in the $B$ and $V$ bands for PZ Mon does not exceed 0.15$^m$. The variability cannot be caused by the cool region, because, in that case, the \bv\ color should not change, but it is variable. This can be explained by the constant presence of a cool region on 41\% of the visible stellar hemisphere, i.e., either the presence of a polar
spot, because at the inclination of the rotation axis of PZ Mon $i=67\deg$ \citep{2015AstL...41..677P} one of the poles is always visible, or a uniform distribution of spots over the entire photosphere. Both options most likely play their role. 

This model describes well the spectral energy distribution for PZ Mon computed, according to \citet{1998A&A...333..231B}, from the fluxes in individual bands (see Fig.~\ref{fig:sed}, where the first point corresponding to the $U$ band was taken from \citet{2018AstL...44...35P}).

\subsection{Warm spot}

The model of a cool polar spot describes well the brightness maxima of PZ Mon, but to describe the amplitudes of brightness variations in different bands, it is necessary to add spots with a temperature higher than that for the polar spot Ts , but lower than that for the quiet photosphere \Teff. The brightness amplitudes can be calculated from a formula similar to (eq.\ref{eq:m}) with the addition of a parameter characterizing warm spots with an area \Swarm

\begin{eqnarray}
\Delta m_X = -2.5\,log [10^{-0.4 \displaystyle m_X^s}\,S_s+10^{-0.4 \displaystyle m_X}\,(1-S_s)]-
\nonumber\\
-2.5\,log[10^{-0.4 \displaystyle m_X^s}\,S_s+10^{-0.4 \displaystyle m_X^w}\,S_w+
\nonumber\\
+10^{-0.4 \displaystyle m_X}\,(1-S_s-S_w)]
\label{eq:mpen}
\end{eqnarray}
\noindent
where the index $s$ refers to the cool spots, $w$ refers to the warm spots, and no index refers to the unspotted surface with \Teff=4730~K. In Eq.(\ref{eq:mpen}) we use the information that the warm spot is located only in the hemisphere on the side of the secondary component, i.e., it does not manifest itself at the brightness maximum of PZ Mon. The polar spot is always present with fixed parameters \Tspot=3500~K and \Sspot=41\%. Since $m_X^w=f(T_w)$, $\Delta m_X=f(T_w,S_w)$ and the temperature and area of the warm spots can be estimated from the observed amplitudes. We have these two unknowns and five Eqs.(\ref{eq:mpen}) for the $BVRIJ$ bands (the accuracy of observations for the longer-wavelength bands is insufficient) and obtain \Twarm=4493$\pm$106~K, \Swarm$=20\pm1$\%\ using the Levenberg--Marquardt method. The computed amplitudes for these parameters describe the observed ones with an accuracy exceeding the observational errors (Fig.~\ref{fig:amp}).

\section{MODEL OF THE SPOTTED PHOTOSPHERE OF PZ Mon}
\label{sec-model}
\noindent

\begin{figure}
	\centering
	\includegraphics[width=\columnwidth,clip]{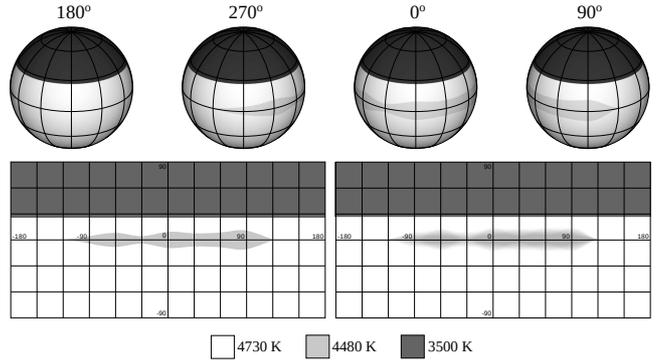}
	\caption{Maps of the temperature distribution over the photosphere of \pz. The left and right maps were constructed based on the computed model parameters and the variations in parameters, respectively.  Orthographic projections of the star at different longitudes of the central meridian are shown at the top.}
	\label{fig:map}
\end{figure}

We constructed a model for the intensity distribution over the surface of the star under study bytechnique similar to that in \citet{2018AstL...44...35P}. On a 720$\times$360 map there are a cool polar spot with a radius of 63.7$\deg$ (this provides 41\% of the visible surface, given the limb-darkening effect) and a warm equatorial region that is represented by a slightly different model in this paper. It is also located along the equator and is symmetric relative to it. The shape of the spot is specified at grid points with a longitudes l from $-90\deg$ to $90\deg$ and a $30\deg$ step. The zero meridian is directed toward the secondary component of this binary system. For all longitudes $|l|>120\deg$ the latitude of the edge of the warm region is set equal to zero, because when observing the opposite hemisphere (this corresponds to the brightness maximum), we describe well this part of the light curve without invoking warm spots. All points were connected by an Akima spline, which gives a smooth shape of the region under consideration, because the values of the spline between the grid points are within the range of their values. As a result, we have eight parameters (seven latitudes and the temperature \Twarm) that we searched for using the Levenberg-Marquardt nonlinear fitting method. The limb-darkening effect was taken into account with the coefficients from \citet{1993AJ....106.2096V}.

We restricted ourselves to fitting the data for only two bands, $B$ and $V$ , because they have the smallest observational error and the most reliable magnitudes. During our modeling we computed the light curves in all bands. As an initial approximation we used \Twarm=4500~K and two types of spots: (1) in the form of a strip?the latitude of all points is the same and the filling factor is equal to the previously found estimate of 20\%; (2) in the form of a wide spot at the central meridian within the range of latitudes $|l|<60\deg$. In both cases, the solutions converged at close values of the parameters and the computed light curves describe the observed ones with a 0.01$^m$ accuracy (Figs.~\ref{fig:BVRI} and \ref{fig:JHKL}). The temperature of the warm spots is \Twarm=4480$\pm$80$\pm$200~K, where the first and second errors correspond, respectively, to the uncertainties in the temperature at fixed spot shape and area and when all parameters are free. 

Figure~\ref{fig:map} presents the modeling results. The left map was constructed based on the computed parameters. The surface temperature is defined by three values: \Teff=4730~K for the quiet photosphere, \Tspot=3500~K for the cool polar spot, and \Twarm=4480~K for the warm equatorial spot. Orthographic projections of the star at different longitudes of the observed central meridian are shown at the top. The right map was constructed based on the variations in parameters. Each parameter, i.e., the latitude of the spot boundary at a specific grid point, was varied by a value from $-4\deg$ to $+4\deg$ with a $2\deg$ step. We performed more than 80\,000 computations of individual maps. Then, each point of the map was averaged with a weight inversely proportional to the square of the error in describing the observations, which ranged from 0.010$^m$ to 0.025$^m$. Due to the averaging, the temperature range for the warm spot \Twarm\ from 4480 to 4720~K is presented on this map. We see that both maps are very similar; an extended region from the zero meridian to $90\deg$ and an almost isolated spot near $l=-60\deg$ are identified. Despite the fact that we used only two of the eight bands in fitting the observations, the computed values for most of the other bands do not differ from the observed ones within the error limits. The largest discrepancies (their values are shown in Fig.~\ref{fig:JHKL} in the lower left corner of each plot) were revealed for the infrared $H$ and $L$ bands. The deviations for these bands are also seen in Figs.~\ref{fig:T} and \ref{fig:dm} and may be related to the observational errors.

\section{DISCUSSION AND CONCLUSIONS}
\label{sec-discussion}
\noindent
The parametric three-component model of the temperature distribution over the photosphere of the active giant PZ Mon constructed in this paper describes all of the light curves in different bands, from the blue $B$ ($\lambda_{cen}=4380$~\AA) to the infrared $L$ ($\lambda_{cen}=34\,500$~\AA), within the observation error limits. The model consists of a quiet photosphere withtemperature \Teff=4730~K, a polar spot with a radius of $63.7\deg$ and a temperature \Tspot=3500~K, and an equatorial region with an intermediate temperature \Twarm=4480~K located in the hemisphere on the side of the secondary component. The map in Fig.~\ref{fig:map} is similar to the map constructed by \citet{2018AstL...44...35P} for the epoch of early 2015, when the light curve exhibited a similar behavior. An extended region to the right of the zero meridian and an isolated spot near $l = -60\deg$ are also present. The entire active warm region is distributed over a wide range of longitudes rather than is concentrated near the zero meridian, as was assumed previously. It seems impossible to identify one single active longitude.

This simple and approximate model does not describe the latitude distribution of spots and the effects of differential rotation, which is also inaccessible to more complex methods based on a light-curve analysis and an inverse-problem solution \citep[see e.g.][]{2008AN....329..364S}) this requires a spectroscopic monitoring of the star and highly accurate photometric data. However, the contribution of both warm and cool spots for each individual longitude must be close to their real values. Doppler tomography and interferometric observations of stars with spots in \rs\ binary systems show a similar situation: both cool polar regions \citep[$\zeta$~And -- ][]{2016Natur.533..217R} \citep[$\sigma^2$~CrB -- ][]{2018A&A...613A..60R} and high-latitude ones \citep[SV~Cam -- ][]{2018MNRAS.479..875S} as well as those distributed over the surface \citep[$\sigma$~Gem -- ][]{2017ApJ...849..120R}. Intermediate temperatures from the coolest ones ($\sim$3500~K) to the temperature of the quiet photosphere are present in all cases. A significant difference between the adopted model and the data from Doppler tomograms and interferometric observations is the polar spot size reaching $20\deg .. 30\deg$, which is considerably smaller than that in our model. The cool region may well be located at both poles. In this case, the radius of the identical polar spots will be $56.4\deg$ , which is only slightly smaller than the size for the spot location only at one pole, because for the adopted inclination of the rotation axis of PZ Mon a considerable area near the opposite pole is invisible. However, even this value is twice that expected from observations. Therefore, we can assume that about half of all cool spots (or an area of 20\%\ of the visible surface) are arranged more or less uniformly outside the poles.

The presence of a significant fraction of cool active regions must be accompanied by the presence of a magnetic field. \citet{2018A&A...613A..60R} mapped the intensity and magnetic field distributions for the system $\sigma^2$~CrB; the magnetic field strength reaches 0.4--0.6~kG in cool active regions. The appearance of magnetic field effects in the stellar spectrum may also be expected for PZ Mon. Previously, based on a spectrum with a resolution $R$ = 40\,000, we found no dependence of the width of spectral lines on their Lande factor, possibly, because of the low spectral resolution (the instrumental line broadening is comparable to the stellar rotation broadening). In future, we are planning to obtain and analyze higher-quality spectra with a resolution of at least 80\,000. The cool regions must manifest themselves better in the red part of the spectrum and there is a possibility to find the traces of a magnetic field.

\section*{ACKNOWLEDGMENTS}
\noindent
S.Yu. Gorda's work was financially supported by the Ministry of Education and Science of the Russian Federation (the basic part of the State contract, RKAAAA-A17-117030310283-7) and the Government of the Russian Federation (resolution 211, contract 02.A03.21.0006).

\bibliographystyle{mnras}
\bibliography{paper}

\end{document}